\documentclass[aps,prl,twocolumn,showpacs,intlimits,amsmath,amssymb,floats]{revtex4-2}
\usepackage{graphicx}
\usepackage{color}
\usepackage{natbib}
\usepackage{amsmath}
\usepackage{amsfonts}
\usepackage{graphicx}
\usepackage{dcolumn}
\usepackage{tabularx}
\usepackage{keyval}
\usepackage{multirow}
\usepackage{xcolor}

\begin{document}
\title{Near-field imaging of domain switching in in-operando VO$_{2}$ devices}
\author{Sergio Salvía Fernández}
\affiliation{van der Waals - Zeeman Institute, University of Amsterdam, Sciencepark 904, 1098 XH Amsterdam, the Netherlands}
\author{Xing Gao}
\affiliation{Faculty of Science and Technology and MESA+ Institute for Nanotechnology, University of Twente, Enschede, the Netherlands}
\author{Silvia Cassanelli}
\affiliation{van der Waals - Zeeman Institute, University of Amsterdam, Sciencepark 904, 1098 XH Amsterdam, the Netherlands}
\author{Stephan Bron}
\affiliation{van der Waals - Zeeman Institute, University of Amsterdam, Sciencepark 904, 1098 XH Amsterdam, the Netherlands}
\author{Hans Hilgenkamp}
\affiliation{Faculty of Science and Technology and MESA+ Institute for Nanotechnology, University of Twente, Enschede, the Netherlands}
\author{Erik van Heumen}
\email{e.vanheumen@uva.nl}
\affiliation{van der Waals - Zeeman Institute, University of Amsterdam, Sciencepark 904, 1098 XH Amsterdam, the Netherlands}
\affiliation{QuSoft, Science Park 123, 1098 XG Amsterdam, The Netherlands}

\begin{abstract}
Experimental insight in the nanoscale dynamics underlying switching in novel memristive devices is limited owing to the scarcity of techniques that can probe the electronic structure of these devices. Scattering scanning near-field optical microscopy is a relatively novel approach to probe the optical response of materials with a spatial resolution well below the diffraction limit. We use this non-invasive tool to demonstrate that it provides detailed information on the origin and memory behaviour of ultra-thin films of vanadium dioxide. Simultaneously recorded $I(V)$ characteristics and near-field maps show that discontinuities in the I(V) characteristics arise from the sudden switching of insulating domains to metallic domains. At the threshold voltage, the domains form a continuous current path. The metallic domains persist once the bias voltage is removed, but narrow monoclinic regions appear around the domain boundaries. The key advantage of our approach is that it provides detailed information on the electronic structure at length scales raging from tens of nanometers up to tens of microns and is easily applied under \textit{in operando} conditions.
\end{abstract}
\date{\today}
\maketitle
Scattering scanning near field optical microscopy (sSNOM) is a versatile, non-invasive probe of the dielectric and plasmonic response of correlated and 2D materials. Based on tapping-mode atomic force microscopy (AFM), an sSNOM can map out optical contrast on a scale of a few tens of nanometers at any wavelength due to the interaction between a polarizable AFM cantilever, an incoming laser and the sample under study. Some of the more recent enhancements of the sSNOM implementation include enabling broadband spectroscopy \cite{huthNL2012, huth2012}, ultra-fast pump-probe microscopy \cite{xu2013, wagner2014, donges2016} and THz sSNOM \cite{stinson2018}. An equally important endeavour has been improving the ability to manipulate material parameters in-situ. Examples include manipulating the optical response of a material through changing the carrier concentration \cite{fei2012,chen2012} or exploring phase transitions with variable temperature microscopy \cite{yang2013,mcleod2017}. Invariably, adding new capabilities to the sSNOM repertoire have led to new insights in the electronic properties of materials on nanometer length scales \cite{liu:RPP2017,chen:AM2019}.      

In this article we map out the near-field optical response \textit{in operando} on a memristive device as it is driven through switching events. The device is based on a vanadium dioxide thin film, which features a reversible insulator to metal transition. Vanadium dioxide is a poster child for the application of complex oxide materials in a variety of room temperature, energy saving applications \cite{zhou:IEEE2015,cui:joule2018}. As platform for a memristor device, VO$_{2}$ has been investigated in a variety of different forms \cite{kumar:AM2013,madan:acsnano2015,delvalle:nature2019,rana:SR2020,delvalle:science2021,nikoo2022,gao:aip2022}. Depending on the details, switching can be as fast as a few 100's of femtoseconds \cite{cavalleri:PRL2001, donges2016}. Particular interest has gone to fully strained films \cite{muraoka:apl2002,kittiwatanakul:APL2014,rodriguez:acsaem2020}. These ultra-thin films are fully in the rutile phase in both the insulating and metal phases, which is expected to lead to enhanced durability of devices.

Several studies have investigated whether the current driven transition arises from electric field effects \cite{gopalakrishnan:jmsi2009,goldflam:apl2014,madan:acsnano2015,rodriguez:acsaem2020,sternbach:NL2021} or from Joule heating \cite{kim:apl2010,zhong:jap2011,kumar:AM2013,zimmers:prl2013}. It appears that both effects are present to some degree and that the balance depends on detailed properties of the structures studied (film thickness, substrate etcetera). In-operando probing of device switching has provided insight in the parameters determining the switching speed \cite{delvalle:science2021}, in the role played by film thickness and crack formation \cite{rodriguez:acsaem2020} and in the role played by nanoscale domains in unstrained films \cite{madan:acsnano2015}. 

Scattering near-field optical microscopy has proven to be a valuable tool for the study of VO$_{2}$ crystals and films \cite{qazilbash:science2007,jones:NL2010,atkin:PRB2012,liu:PRB2013,liu:APL2014,ocallahan:NC2015}, but to the best of our knowledge has not been applied to the study of in-operando devices or fully strained thin films. In this article we use it to compare the insulator to metal transition (IMT) both as function of temperature and applied electric field with 20 nm spatial resolution. We investigate the heating induced by the CO$_{2}$ laser and its effect on the onset of the MIT. We use the sSNOM contrast to elucidate the mesoscopic changes in the electronic structure as a current is applied. Combined with simultaneously measured $I(V)$ characteristics, this enable us to pinpoint the origin of irregular steps in $I(V)$ traces.

\begin{figure}[h]
    \includegraphics[width=\columnwidth]{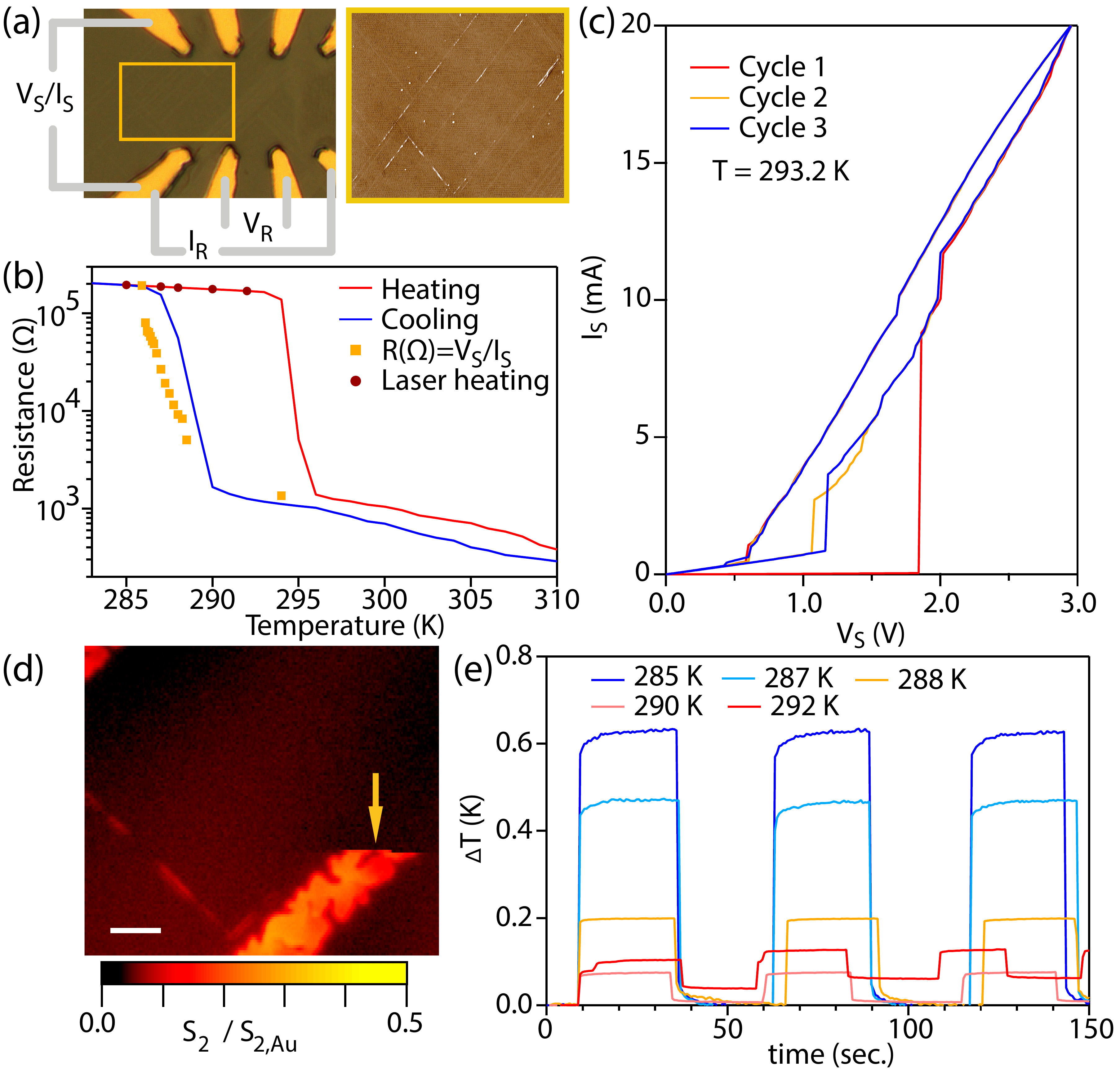}
    \caption{(a) Experimental geometry. The yellow square indicates the relation between AFM topography (on the right) and the contacts. (b): Resistance ($R\,=\,V_{R}/I_{R}$) recorded during heating and cooling cycle. Also indicated are the two-point resistances $V_{S}/I_{S}$, measured during the sSNOM experiments (yellow squares). A small shift in temperature is due to laser-induced heating. Red solid circles indicate temperatures where we investigated heating due to the CO$_{2}$ laser, see panel (e). (c): $I_{S}(V_{S})$ characteristic measured at 293.2\,K. In the first cycle the resistance is high until a threshold voltage is reached at 1.7 V. The $I_{S}(V_{S}$) characteristic then follows a different voltage dependence until the sample is reset by cooling well below the IMT. (d) sSNOM image recorded at 292\,K during a heating cycle. The arrow indicates the sudden appearance of a metallic domain induced by laser heating. (d): change in local temperature determined from the corresponding change in resistance when the CO$_{2}$ laser is focused on the sample. See the text for more details.}
    \label{Fig1}
\end{figure}

In our experiments, we combine scattering scanning near-field optical microscopy (sSNOM) with transport measurements. Strained VO$_2$ films of 13 nm thickness are grown on a single crystal TiO$_2$ substrate using pulsed laser deposition \cite{rana:SR2020}. Subsequently, sets of eight gold contacts are created using photolithography (Fig. \ref{Fig1}a). Each set consists of two rows of four contacts that are separated by 3\,$\mu \text{m}$, while the rows are separated by 12\,$\mu\text{m}$. This layout allows us to determine the two-point resistance from $V_{S}/I_{S}$ by using pairs of contacts or a four-point resistance, $R\,=\,V_{R}/I_{R}$, by employing the four contacts on a row. The smaller spacing between contacts on a row is less ideal for the sSNOM experiments, as only a small region between the contacts can be probed. We therefore use two-point $I_{S}(V_{S})$ measurements between a contact in the top and bottom row to simultaneously record the resistance, while scanning the near-field response over a large 15\,$\mu\text{m}$ by 10\,$\mu\text{m}$ area. We choose this area to be in between the contacts (yellow square in Fig. \ref{Fig1}a) over which a voltage $V_{S}$ is applied, so that the current flows approximately perpendicular to the scanning direction. We use the gold contacts as reference for the near-field maps, enabling quantitative comparisons between experiments and eliminating small changes in the CO$_{2}$ laser intensity. All sSNOM data is displayed on a color scale labelled as S$_{2}$/S$_{2,Au}$. In other words, the value of the second harmonic of the demodulated near-field signal in every pixel, is divided by the average value of the corresponding signal measured on a nearby gold contact. Such gold reference images are recorded frequently, e.g. after changing temperature or voltage. AFM topographies are collected as well, showing an atomically smooth surface over the entire field of view with only a few domain lines across the field of view (white, diagonal lines in Fig. \ref{Fig1}a). The near-field optical response is measured at a wavelength $\lambda\,=\,10.7\,\mu\text{m}$ (116 meV). At this wavelength large optical contrast is expected between metallic and insulating regions \cite{qazilbash:science2007}. 
X-ray diffraction measurements on the as-deposited VO$_{2}$ film indicate that the film was principally in rutile phase. The associated strain is accommodated by monoclinic-phase planar defects representing the domain lines mentioned above. As observed previously, films with thickness below 35 nm are almost completely in the rutile phase \cite{kittiwatanakul:APL2014,rodriguez:acsaem2020}. The insulator-to-metal transition (IMT) in our films is therefore purely electronic in nature. Figure \ref{Fig1}b shows that the resistance changes by almost three orders of magnitude with transition temperatures during $T_{C}$\,=\,289\,K and $T_{H}$\,=\,295\,K, recorded during cooling (C) and heating (H) respectively. The width of the transition is approximately 2 K, with a hysteresis of 6 K. 

The application of a voltage while in the insulating state can also trigger the IMT \cite{berglund:ieee1969,kim:apl2010,zhong:jap2011,rodriguez:acsaem2020,rana:SR2020}, as shown in Fig. \ref{Fig1}c. Starting from a temperature below the IMT (293 K in Fig. \ref{Fig1}c), the system is in a high resistance state until a threshold voltage is reached. Beyond this threshold the current increases hundredfold, signalling the formation of a metallic state. We will call this the forming cycle. In subsequent $I(V)$ cycles the metallic state persists, as evidenced by the distinctly different cycles 2 and 3 in Fig. \ref{Fig1}c. In these subsequent cycles small steps in current at particular voltages are also present, which are close to reproducible from cycle to cycle. The insulating state can be restored by cooling below $T_{C}$.

\begin{figure*}[ht]
    \includegraphics[width=\textwidth]{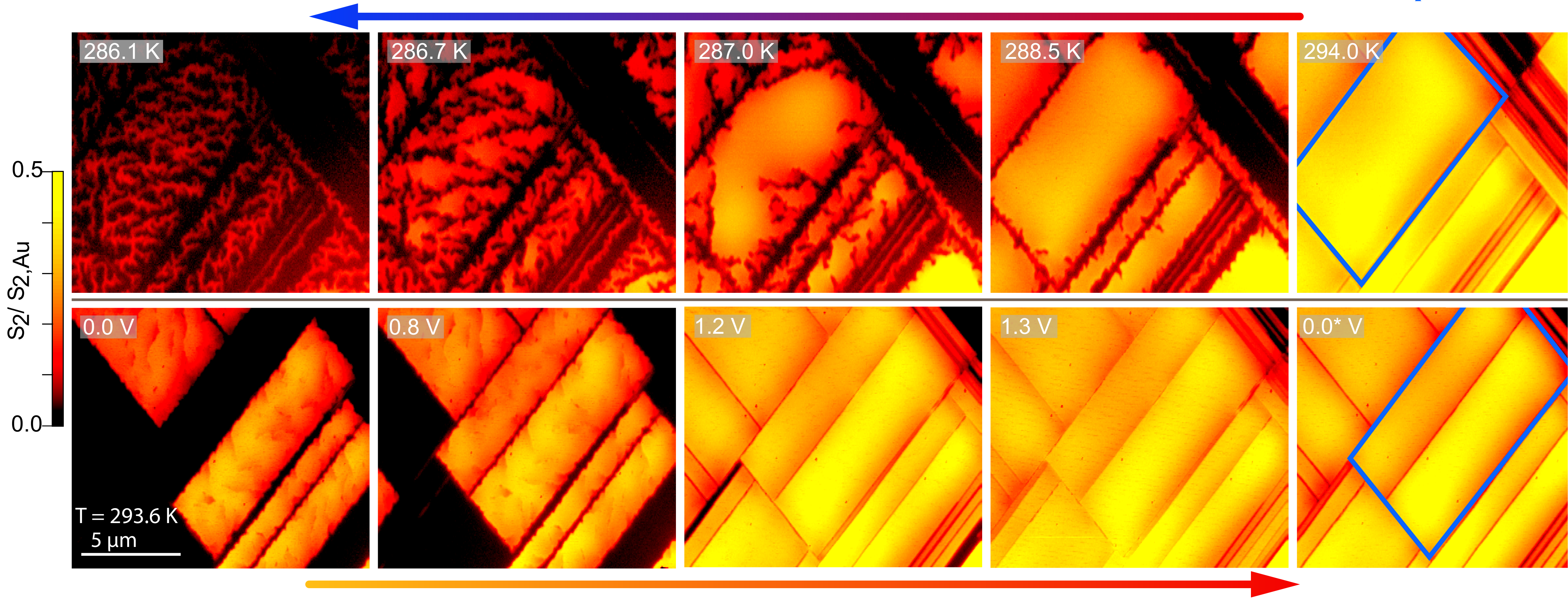}
    \caption{Top row: Gold referenced near-field scattering amplitude $S_{2}/S_{2,Au}$ maps recorded at selected temperatures while cooling from the metallic state. Bottom row: $S_{2}/S_{2,Au}$ recorded at various voltages V$_\text{S}$ during a forming $I(V)$ cycle (at 293.2\,K). The threshold voltage for this measurement occurred at 1.25 V. The panel labelled with 0.0* V was recorded immediately after turning off the applied voltage. Blue squares in the two right-most panels indicate the same domain. }
\label{Fig2}
\end{figure*}

One of the interesting aspects of VO$_{2}$ is that, besides dc electric fields, the IMT can be induced through other external knobs such as strain \cite{atkin:PRB2012} and ultrafast light pulses \cite{lysenko:ass2006,sternbach:NL2021}. This also plays an important role in our experiments as demonstrated in Fig. \ref{Fig1}d. The sSNOM data is taken at T\,=\,292 K close to $T_{H}$ in the insulating state. Indeed, a large part of the field of view (FOV) has a low near-field amplitude, indicative of the insulating state \cite{qazilbash:science2007}. In the middle of the scan we observe that the CO$_{2}$ laser used in the experiment is also capable of inducing the IMT. This can be seen by the sudden switching of a domain (indicated by a yellow arrow). Our experimental geometry allows us to determine the laser-induced heating by effectively using the induced change in resistivity as thermometer. 

For this, we start by cooling the sample to 283\,K and then slowly heat the film to a selected temperature. With the AFM engaged, we turn the CO$_{2}$ laser `on' and `off' with a flip mounted mirror. The effective temperature increase in the film can now be estimated by monitoring the change in resistivity that is measured in the dark and illuminated state. To convert resistivity to temperature readings, we invert the resistivity data of Fig. \ref{Fig1}b at the temperatures shown by dark-red solid circles.  Figure \ref{Fig1}e shows the resulting $\Delta T$ traces as function of time. Depending on the starting temperature the effect of the laser is to increase the temperature of the film by as much as 0.6 K for the lowest measured temperature. The magnitude of $\Delta T$ decreases with increasing starting temperature. This is likely related to the changes in absorption coefficient, specific heat and thermal conductivity that take place near the IMT. We notice that close to $T_{H}$ the resistivity does not return to the original value, resulting in a finite $\Delta T$ also in the `off' state. This can be related to the switching of a large domain to the metallic state induced by the laser, similar to what is observed in Fig. \ref{Fig1}d. As we will follow the IMT using sSNOM and would like to exclude sudden switching resulting from laser heating, we carry out all subsequent experiments while cooling from the metallic state. For every measurement, we set the temperature and record the two-point resistance between contacts at the top and bottom of the image. The $V_{S}/I_{S}$ values obtained in this way are shown in Fig. \ref{Fig1}c as yellow squares. Note that they cannot be directly compared to the four-point resistance (solid lines) since the two-point resistance probes a different area of the surface.

Figure \ref{Fig2} shows the main experimental result of this paper: sSNOM images recorded as function of temperature (top row) and as function of applied voltage $V_{S}$ (bottom row). We first focus on the evolution of the near-field signal, recorded while cooling down from 343\,K. Initially, the near-field response is fully metallic, but near the phase transition, some domains start to become insulating. As has been observed previously \cite{rodriguez:acsaem2020}, thin films often consist of domains that can have slightly different critical temperatures. In films with a thickness below 15 nm, the domains can become very large and even occupy the whole film.

The 294 K panel shows an almost completely metallic film (corresponding to yellow in our color scale), with thin lines (dark) that follow the white domain lines observed in the topography of Fig. \ref{Fig1}a. We note again that these lines are thought to be VO$_{2}$ in a monoclinic phase. At 288.5\,K meandering, insulating filaments emerge that creep in from the domain boundary. These filaments permeate the domain as temperature is further decreased. One novel aspect compared to previous experiments \cite{qazilbash:science2007,jones:NL2010,atkin:PRB2012,liu:PRL2013,liu:PRB2015,ocallahan:NC2015} is that our films consist of large single crystalline domains that are fully strained and thus already in the rutile phase. The transition within each domain is therefore purely electronic and this reflects itself in only weak pinning of the nucleation sites of the filaments. As the experiment is repeated, the formation of filaments starts from different points on the boundary.

In the bottom row panels of Fig. \ref{Fig2} we contrast the temperature dependence of the near-field signal with the voltage dependence of the signal. As starting point we first cool the film to 283\,K and subsequently heat the film to 293.2\,K, two degrees below $T_{H}$. Without applying any voltage, we observe that some domains are already metallic (panel labelled 0.0\,V). Comparing the $V_{S}$\,=\,0\,V data with the data presented in the top row of Fig. \ref{Fig2}, we note that the formation of the metallic state is reversed as compared to the formation of the insulating state. One can still see insulating filaments within the domain that are absorbed in the metallic matrix as the voltage is increased. As we increase the voltage from $V_{S}$\,=\,0\,V to $V_{S}$\,=\,0.8\,V, we observe two small steps in the current that flows between the contacts. This change in current appears to correspond to the switching of two complete domains: one in the middle of our field of view and one on the far left side of it. Time domain reflectivity experiments suggest that the formation of filaments within a domain can take place on microsecond time scales \cite{delvalle:science2021}, which is too fast for our scanning method to register and we consequently only see the end state in which the domain is fully metallic.

While performing the sSNOM measurements, we noticed a large step in current is observed at $V_{S}$\,=\,1.15\,V, at a lower threshold voltage than the step seen in Fig. \ref{Fig1}c. This offset in threshold voltage is due to the additional heating induced by the CO$_{2}$ laser as can be verified by repeating the experiment with the laser in the `off' state. In the $V_{S}$\,=\,1.3\,V data presented in Fig. \ref{Fig2} we see that all domains in our field of view are fully in the metallic state, with the exception of a few narrow domains in the top right corner. We also note that the average intensity of the domains increases and that insulating filaments within a domain have disappeared. 

Unexpectedly, only a small change is observed when the voltage is set to zero, $V_{S}$\,=\,0*\,V. The most notable change is that the lines between domains become more prominent and increase in width. However, the average sSNOM intensity within domains hardly decreases. We monitored the persistence of this state over time using a small probing current (1\,$\mu $A). We observed no significant change in the resistivity over a 17 hour period in this way. We have therefore clearly identified the source for the forming effect in the first cycle: starting from a mostly insulating phase 'low' state (0.0 V), metallic domains form upon application of a voltage (0.8 V). As the voltage increases, more domains become conducting and the device changes to a 'high' state when all domains between the contacts are in the metallic state (1.2\,-\,1.3 V). As the device is turned off, the interior of the domains remain metallic while the domains are isolated from each other through narrow insulating regions. When we go through a second cycle, these insulating regions will again become metallic giving rise to the small steps seen in cycle 2 and 3 in Fig. \ref{Fig1}c).  
\begin{figure*}[th]
    \includegraphics[width=\textwidth]{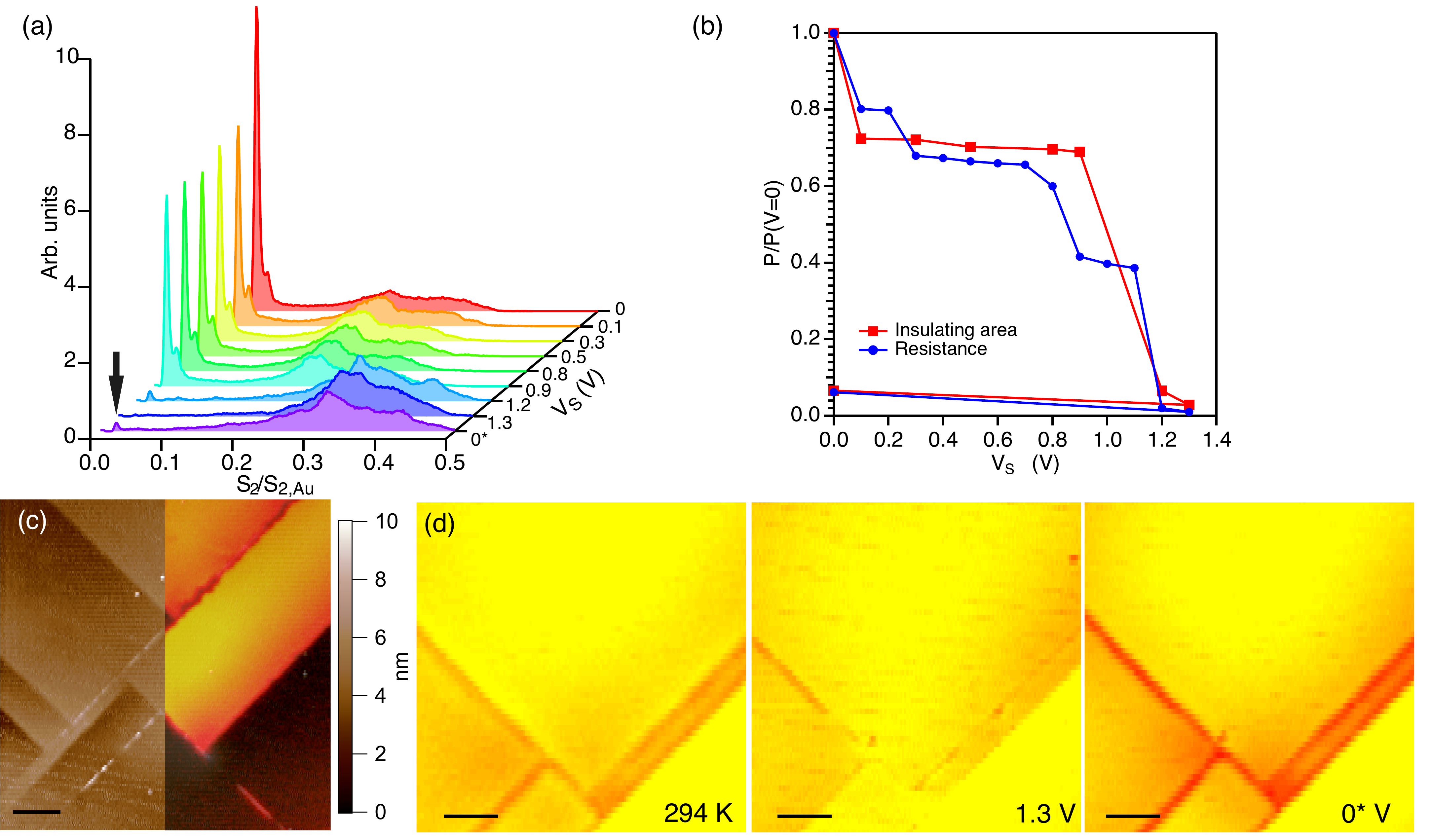}
    \caption{(a): histograms corresponding to the sSNOM images recorded during $I(V)$ cycles. The signal below 0.1\,V corresponds to insulating domains. At 1.2 V there is a notable shift in the position of the high signal maximum. When the voltage is set to zero, a small insulating component reappears, indicated by the arrow. (b): Comparison of the relative changes $P(V_{S})/P(V\,=\,0)$ of the resistance and insulating area. (c): topography measured at 5 V and 292 K partly overlayed with S$_{2}$/S$_{2,Au}$. The large height differences (8 nm) are not observed at lower voltages. Scale bar corresponds to 1$\mu$m. (d): comparison of monoclinic areas around domain lines observed in topography at 294 K, at $V_{S}$\,=\,1.3\,V and $V_{S}$\,=\,0*\,V. The topography does not change at $V_{S}$\,=\,1.3\,V, but the surrounding area noticeably undergoes an IMT. Scale bars correspond to 500 nm.}
\label{Fig3}
\end{figure*}

A more quantitative picture of the changes underlying the $I(V)$ characteristics can be obtained by plotting the intensity variation within a field of view in a histogram representation. Fig. \ref{Fig3}a shows the histograms for the (extended) series of measurements presented in the bottom row of Fig. \ref{Fig2}. At $V_{S}$\,=\,0\,V, most of the domains are insulating and this is represented by a large, sharp peak in the histogram with average signal intensity S$_{2}$/S$_{2,Au}$\,$\approx$\,0.02. A broader peak centred around S$_{2}$/S$_{2,Au}$\,$\approx$\,0.25 can also be seen and corresponds to the metallic domains. As $V_{S}$ is increased, some weight shifts from the  S$_{2}$/S$_{2,Au}$\,$\approx$\,0.02 peak to S$_{2}$/S$_{2,Au}$\,$\approx$\,0.25 until the threshold voltage $V_{S}$=1.15 is reached. As can be seen from the histogram, at $V_{S}$\,=\,1.2\,V the maximum around S$_{2}$/S$_{2,Au}$\,$\approx$\,0.25 suddenly shifts to higher signal values (S$_{2}$/S$_{2,Au}$\,=\,0.35) and the sharp peak at S$_{2}$/S$_{2,Au}$\,=\,0.02 is strongly reduced. Finally at $V_{S}$\,=\,1.3\,V the remaining insulating regions around the domain lines turn metallic and the S$_{2}$/S$_{2,Au}$\,$\approx$\,0.02 peak disappears. When the voltage is set to $V_{S}$\,=\,0, a small signal at S$_{2}$/S$_{2,Au}$\,$\approx$\,0.02 reappears, signalling a small contribution from the insulating phase.

The mesoscopic transport behaviour of the film can be largely predicted by the changes taking place in the histograms. Figure \ref{Fig3}b compares relative changes in the insulating area at a particular voltage to the insulating area at $V_{S}$\,=\,0\,V (e.g. $A_{ins.}(V_{S})/A_{ins.}(V_{S}\,=\,0))$, with the relative change in resistance ($R(V_{S})/R(V_{S}\,=\,0$). To obtain the relative change in insulating area, $A_{ins.}(V_{S})$, we integrate the histograms of panel \ref{Fig3}a up to S$_{2}$/S$_{Au}$\,=\,0.1. As Fig. \ref{Fig3}b shows, the changes in resistance are accompanied by distinct changes in the total area of insulating domains. Small differences (for example, the step in resistance around $V_{S}$\,=\,0.8\,V) are due to the fact that some domains that contribute to the total current path fall outside our field of view. 

Structural effects have little influence on the mesoscopic properties of the film, as demonstrated in Fig. \ref{Fig3}c. Here we show a partial overlay of near-field (right half) on the topography (left half) under the application of $V_{S}$\,=\,5\,V at 292.5\,K. At such large applied voltages, which were applied only as our final measurements, it is sometimes seen that the film partly detaches from the underlying substrate. Despite the 8 nm height variations between some domains, we observed only a weak impact on the resistance or the near-field signal. The dominant effect underlying the memristive device in our experiments, is therefore consistent with Joule heating \cite{kim:apl2010,zhong:jap2011,kumar:AM2013,zimmers:prl2013}. In contrast to earlier work \cite{kumar:AM2013}, we find that the current flow between contacts selectively heats domains above their transition temperature. This points to small variations of $T_{H}$ in individual domains as has been observed previously \cite{rodriguez:acsaem2020}. Although we cannot completely exclude electric field effects, we note that reversing the sign of the voltage resulted in the same overall behaviour of the current and near-field signal. 

It has been pointed out previously that strain release in thicker films will lead to domain formation and that domain boundaries can become monoclinic or will have lower strain as compared to the larger domain areas \cite{rodriguez:acsaem2020}. Our direct measurement of the local optical response shows that these regions are quite sharp (Fig. \ref{Fig3}d). Above the IMT, the data taken at 294 K shows large metallic domains separated by domain lines from which a narrow insulating region extends. The spatial resolution in our experiments is approximately 20 nm, enabling us to determine that the insulating regions along the domain lines are 100 - 300 nm wide. Interestingly, the physical domain lines observed in topography are much sharper than this as Fig.'s \ref{Fig1}a and \ref{Fig3}c demonstrate. As temperature is increased, these regions eventually turn metallic as well. In areas of the film with a large domain line density, the intermediate area has a distinctly higher critical temperature as can be clearly seen in the 294\,K data presented in Fig. \ref{Fig2} (top right corner). Similarly, the small, reproducible steps in cycle 2 and cycle 3 (Fig. \ref{Fig3}d), are a result of the switching of insulating domain lines separating the large metallic domains. To conclude, we note that the near-field signal near the insulating strips is a bit lower, indicating perhaps weak effects of reduced strain within the domain itself.


To summarise, nanoscale near-field imaging of \textit{in operando} devices is a feasible and promising new tool to probe the connection between microscopic electronic properties and mesoscopic transport. When applied to rutile VO$_{2}$ devices, it shows that a memory effect is created by persistent switching of large domains. Our 20 nm spatial resolution allows us to demonstrate that $I(V)$ characteristic can be accurately explained from the behaviour of 100 - 300 nm wide insulating strips that are connected to domain boundaries in the film. 

\bibliography{VO2_bibfile}
\end{document}